\documentclass[a4paper,twocolumn]{esapub2005} 
\usepackage{natbib, graphicx, fancyheadings,microtype}
\usepackage{txfonts}

\title{\vspace*{-2.2cm}{\Large\sf\bfseries Tracking of supergranules -- Does it make any sense?}}
\author[, 1, 2]{M. \v{S}vanda \thanks{\noindent\it Cooperating with J.~Zhao, A.~G.~Kosovichev, and Th.~Roudier}}
\author[1]{M. Klva\v{n}a}
\author[1]{M. Sobotka}
\affil[1]{\it Astronomical Institute (v.~v.~i.), Academy of Sciences, Fri\v{c}ova 298, CZ-251 65, Ond\v{r}ejov, Czech Republic}
\affil[2]{\it Astronomical Institute, Charles University, V Hole\v{s}ovi\v{c}k\'ach 8, CZ-180 00, Prague, Czech Republic}
\affil[$\star$]{\it Emails: michal@astronomie.cz, mklvana@asu.cas.cz, msobotka@asu.cas.cz. }

\newcommand{\arcsec}{$^{\prime\prime}$}
\newcommand{\mps}{m\,s$^{-1}$}
\newcommand{\degr}{$^\circ$}

\begin{document}
\clubpenalty=10000
\widowpenalty=10000

\newcommand\apj{{ApJ}}%
\newcommand\apjl{{ApJ}}%
\newcommand\ao{{Appl.~Opt.}}%
\newcommand\aap{{A\&A}}%
\newcommand\solphys{{Sol.~Phys.}}%
\newcommand\nat{{Nature}}%
\newcommand\mnras{{MNRAS}}%

\renewcommand{\headrulewidth}{0pt}
\lfoot{\sf 12$^{\mathsf{th}}$ European Solar Physics Meeting, 8--12 Sep 2008, Freiburg, Germany}
\cfoot{  }
\rfoot{\sf \thepage}


\maketitle
\thispagestyle{empty}

\begin{abstract}
The motions of the plasma and structures in and below the solar photosphere is not well understood. The results obtained using various methods cannot be in general considered as consistent, especially in details. In this contribution we show a summary of the results obtained by the method we have developed recently.

To study the photospheric dynamics we apply the local correlation tracking algorithm to the series of full-disc Dopplergrams obtained by Michelson Doppler Imager (MDI) aboard the SOHO satelite. The dominant structure recorded in Dopplergrams is the supergranulation. Under the assumtion that the supergranules are carried by the flow field of the larger scale, we study properties of this underlying velocity field. The methodology consists of an extensive data processing of primary data in order to suppress disturbing effects such as $p$-modes of solar oscillations or instrumental issues. Aditional coordinate transformations are also needed to make the data suitable for tracking.

We perform comparative tests with synthetic data with known properties and with results of the time-distance helioseismology with a great success. Correlation coeficients of the comparison of mean components of the flow field are larger than 0.8, for the comparison of details in the vector velocity field the correlation coeficient is larger than 0.6. The results of the method applied to real data agree well with well-known features detected in the photospheric velocity fields and reported by many other authors. A few case studies are shown to demonstrate the performance of the method.

As a conclusion let's answer the question in the title. We believe that tracking of supergranules makes a perfect sense when studying the large-scale flows in the solar photosphere. The method we demonstrate is suitable to detect large-scale velocity field with effective resolution of 60\arcsec{} and random error of 15~\mps. We believe that our method may provide a powerful tool for studies related to the dynamic behaviour of plasmas in the solar photosphere. 
\end{abstract}

\section{Introduction}
The solar photosphere is a very dynamic layer of the solar atmosphere. It is strongly influenced by the underlying convection zone. Despite years of intensive studies, the velocity fields in the solar photosphere remain not very well known. The evidence of the vigorous and sometimes chaotic character of the motions of observed structures in the photosphere (sunspots, granules, and other features) came already from the first systematic studies made in 19th century, of which let us at least mention the discovery of the solar differential rotation \citep{1859MNRAS..19...81C}. Motions in the photosphere are strongly coupled with magnetic fields. The large-scale velocity fields are very important for studies of the global solar dynamo. 

An attempt to describe the differential rotation by a parabolic dependence did not make for clear results. Coefficients of the parabola differed with traced objects and also changed with time, when tracing one type of object \citep[reviewed e.\,g. by][]{1985SoPh..100..141S}. The character of the differential rotation has never been in doubt.

It follows from these arguments that a temporally variable streaming of the plasma exists on the surface of the Sun, which can be roughly described by the differential rotation. This streaming has a large-scale character, large-scale plasma motions were studied for example on the basis of tracking the magnetic structures \citep[e.\,g.][]{2001SoPh..198..253A,2001SoPh..199..251A}. The long-term Doppler measurements done by the MDI onboard SOHO make it possible to extend the studies of large-scale velocities in the solar photosphere. The knowledge of the behaviour of velocities in various periods of the solar activity cycle could contribute to understanding the coupling between the velocity and magnetic fields and of the solar dynamo function.

There are at least three methods calculating photospheric velocities:

\begin{enumerate}
\item \emph{Direct Doppler measurement} -- provides only one component (line-of-sight) of the velocity vector. These velocities are generated by local photospheric structures, amplitudes of which are significantly greater than amplitudes of the large-scale velocities. The complex topology of such structures complicates an utilisation for our purpose. Analysing this component in different parts of the solar disc led to very important discoveries \citep[e.\,g. supergranulation --][]{1954MNRAS.114...17H,1962ApJ...135..474L}.

\item \emph{Tracer-type measurement} -- provides two components of the velocity vector. When tracing some photospheric tracers, we can compute the local horizontal velocity vectors in the solar photosphere. Tracking motions of sunspots across the solar disc led to the discovery of the differential rotation \citep{1859MNRAS..19...81C}.

\item \emph{Local helioseismology} -- provides a full velocity vector. Although local helioseismology \citep[e.\,g.][]{2001ApJ...557..384Z} is a very promising method, it is still in progress and until now does not provide enough reliable results.
\end{enumerate}

Since the photosphere is a very thin layer (0.04~\% of the solar radius), the large-scale photospheric velocity fields have to be almost horizontal. Then, the tracer-type measurement should be sufficient for mapping the behaviour of such velocities. In this field the local correlation tracking (LCT) method is very useful.

\begin{figure*}
\centering
\includegraphics[width=0.7\textwidth]{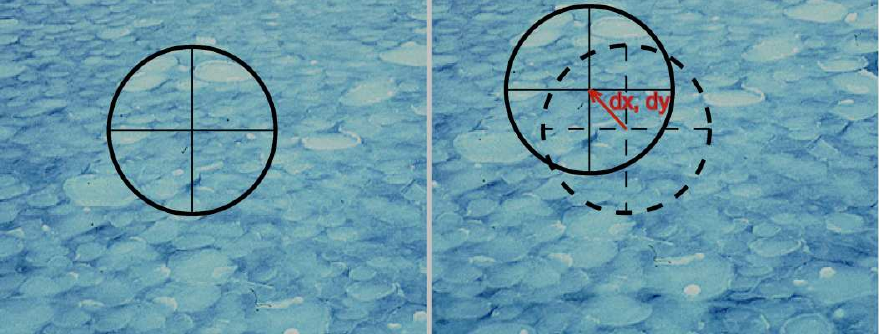}
\caption{The sketch showing the performance of LCT method in detection of corresponding subframes in two frames separated by certain time-lag.}
\label{fig:lct}
\end{figure*}

This method was originally designed for the removal of the seeing-induced distortions in image sequences \citep{1986ApOpt..25..392N} and later used for mapping the motions of granules in the series of white-light images \citep{1988ApJ...333..427N}. The method works on the principle of the best match of two frames that record the tracked structures at two different instants. For each pixel in the first frame, a small correlation window is chosen and is compared with a somewhat displaced window of the same size in the second frame. A vector of displacement is then defined as a difference in the coordinates of the centres of both windows when the best match is found (cf. Fig.~\ref{fig:lct}). The velocity vector is calculated from this displacement and the time lag between two frames.

The method needs a tracer -- a significant structure recorded in both frames, the lifetime of which is much longer than the time lag between the correlated frames. We decided to use the supergranulation pattern in the full-disc Dopplergrams, acquired by the MDI onboard SOHO. We assume that supergranules are carried as objects by the large-scale velocity field. This velocity field is probably located beneath the photosphere, so that the resulting velocities will describe the dynamics in both the photospheric and subphotospheric layer. The existence of the supergranulation on almost the entire solar disc (in contrast to magnetic structures) and its large temporal stability make the supergranulation an excellent tracer. 

The resulting velocities cover the whole solar disc. Hence the velocity field also describes a motion of the supergranulation in the areas occupied by active regions or by magnetic field concentrations. This fact allows the results to be used to study the mutual motions of substructures like sunspots, magnetic field in active regions, background fields, and the quiet photosphere.

The supergranules are structures with a strong convection coupling. The mean size of the supergranular cell is approx. 30~Mm, with the size and shape that may dependent on the phase of the solar cycle. Supergranules are quite stable with a mean lifetime of approx. 20~hours. The internal velocity field in the supergranular cell is predominantly horizontal with the amplitude approx. 300~m\,s$^{-1}$. Due to the horizontality of the internal velocity field, the supergranules can be observed in Dopplergrams on the whole solar disc except for its centre.

We do not propose that this method is capable of measuring velocities of order 1~m\,s$^{-1}$, but we do expect that the large-scale velocities will have magnitudes at least one order greater. We also have to take the largest-scale velocities into account like the differential rotation or meridional circulation, which have velocities of at least 10~m\,s$^{-1}$. If for example we take differential rotation described by the formula $\omega {\rm[deg/day]}= 13.064-1.69 \sin^2 b-2.35 \sin^4 b$, we have to expect a velocity approx. 190~m\,s$^{-1}$ in $b=60\,^\circ$ in the Carrington coordinate system. The main goal of our study (and the proposed method of mapping the velocities should be proxy for this purpose) is to separate the superposed velocity field into the components and to investigate their physics.

\section{Tunning the method}

A recent experience with applying this method to observed data \cite{svanda05} has shown that for the proper setting of the parameters and for the tuning of the method, synthetic (model) data with known properties are needed. The synthetic data for the analysis come from a simple simulation (SISOID code = \emph{SI}mulated \emph{S}upergranulation as \emph{O}bserved \emph{I}n \emph{D}opplergrams) with the help of which we can reproduce the supergranulation pattern in full-disc Dopplergrams.

\begin{figure}[!b]
\includegraphics[width=0.5\textwidth]{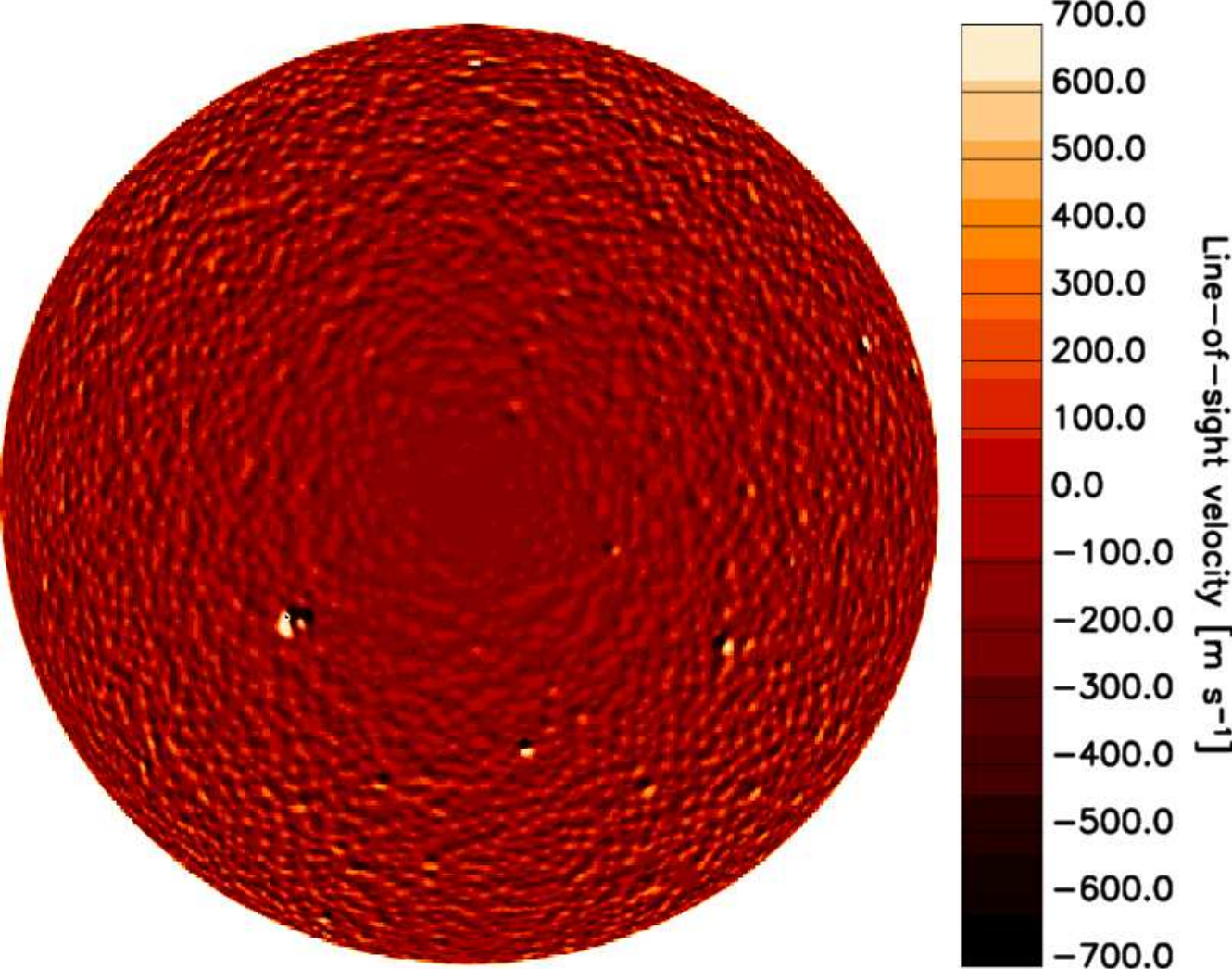}
\caption{An example of the synthetic full-disc Dopplergram produced by SISOID code.}
\label{fig:synthetic}
\end{figure}

The SISOID code is not based on physical principles taking place in the origin and evolution of supergranulation, but instead on a reproduction of known parameters that describe the supergranulation. Individual synthetic supergranules are characterised as centrally symmetric features described by their position, lifetime (randomly selected according to the measured distribution function of the supergranular lifetime), maximal diameter (randomly according to its distribution function) and characteristic values of their internal horizontal and vertical velocity components (randomly according to their distribution function). For details about the SISOID code see \cite{2006AA...458..301S}. An example of the calculated synthetic full-disc Dopplergram can be seen in Fig.~\ref{fig:synthetic}.

\section{Method of data processing}

\begin{figure*}
\includegraphics[width=0.95\textwidth]{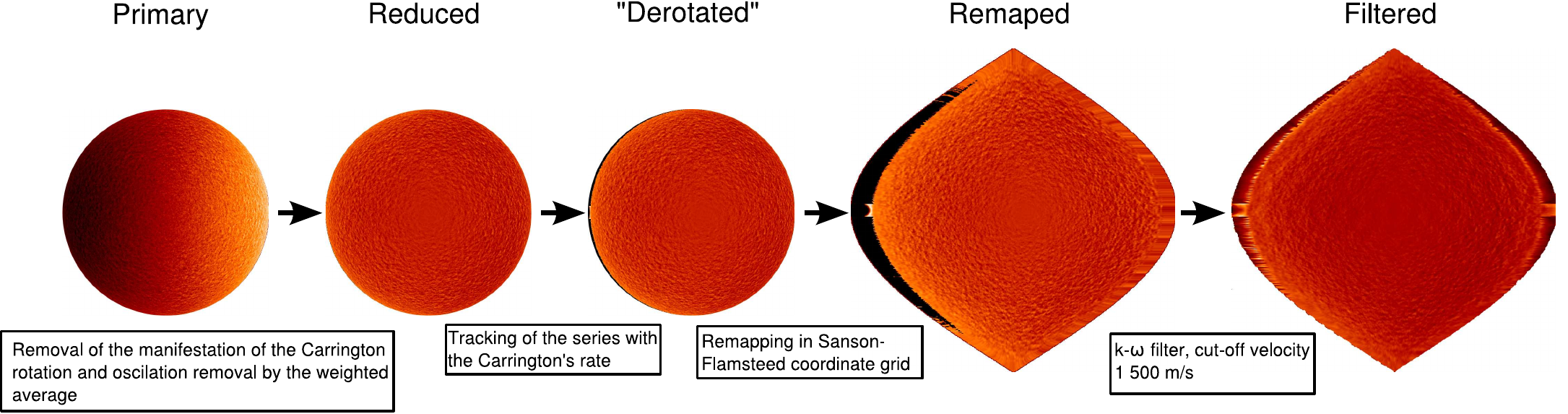}
\caption{Cartoon showing the individual steps in the data processing.}
\label{fig:sketch}
\end{figure*}

The MDI onboard SOHO acquired the full-disc Dopplergrams at a high cadence in certain periods of its operation  -- one observation per minute. These campaigns were originally designed for studying the high-frequency oscillations. The primary data contain lots of disturbing effects that have to be removed before ongoing processing: the rotation line-of-sight profile, $p$-modes of solar oscillations. We detected some instrumental effects connected to the data-tranfer errors. It is also known that the calibration of the MDI Dopplergrams is not optimal and has to be corrected to avoid systematic errors. While examining long-term series of MDI Dopplergrams, we have met systematic errors connected to the retuning of the interferometer. We should also take those geometrical effects into account (finite observing distance of the Sun, etc.) causing bias in velocity determination. In Fig.~\ref{fig:sketch} the cartoon showing the individual steps in the data processing is provided.

As input to the data processing we take a one-day observation that contains 96 full-disc Dopplergrams in 15-minute sampling. Structures in these Dopplergrams are shifted with respect to each other by the rotation of the Sun and by the velocity field under study. 

First, the shift caused by the rotation has to be removed. For this reason, the whole data series (96 frames) is tracked with Carrington rotation rate, so that the heliographic longitude of the central meridian is equal in all frames and also equals the heliographic longitude of the central meridian of the central frame of the series. This data-processing step causes the central disc area (``blind spot'' caused by prevailing horizontal velocity component in supergranules) in the derotated series to move with the Carrington rate. During the ``derotation'' the seasonal tilt of the rotation axis towards the observer (given by $b_0$ -- heliographic latitude of the centre of the disc) is also removed, so that $b_0=0$ in all frames. 

Then the data series is transformed to the Sanson-Flamsteed coordinate system to remove the geometrical distortions caused by the projection of the sphere to the disc. Parallels in the Sanson-Flamsteed pseudocylindrical coordinate system are equispaced and projected at their true length, which makes it an equal area projection.

The noise coming from the evolutionary changes in the shape of individual supergranules and the motion of the ``blind spot'' in the data series with the Carrington's rotational rate are suppressed by the $k$-$\omega$ filter in the Fourier domain. The cut-off velocity is set to 1\,500~m\,s$^{-1}$ and has been chosen on the basis of empirical experience.

The existence of the differential rotation complicates the tracking of the large-scale velocity field, because the amplitudes and directions of velocities of the processed velocity field have a significant dispersion. We have found that, when the scatter of magnitudes is too large, velocities of several hundred m\,s$^{-1}$ cannot be measured precisely by the LCT algorithm where the displacement limit for correlation was set to detect velocities of several tens of m\,s$^{-1}$. Therefore the final velocities are computed in two steps. The first step provides a rough information about the average zonal flows using the differential rotation curve
\begin{equation}
\omega = A+B\sin^2 b+C\sin^4 b\ ,
\label{svanda_eq:fay}
\end{equation}
and calculating its coefficients.

In the second step this average zonal flow is removed from the data series, so that during the ``derotation'' of the whole series the differential rotation inferred in the first step and expressed by~(\ref{svanda_eq:fay}) is used instead of the Carrington rotation. The scatter of the magnitudes of the motions of supergranules in the data transformed this way is much smaller, and a more sensitive and precise tracking procedure can be used.

The LCT method is used in both steps. In the first step, the probed range of velocity magnitudes is set to 200~m\,s$^{-1}$, but the accuracy of the calculated velocities is roughly 40~m\,s$^{-1}$. In the second step the range is only 100~m\,s$^{-1}$ with much better accuracy. The lag between correlated frames equals in both cases 16 frame intervals (i.\,e. 4 hours in solar time), and the correlation window with FWHM 30~pixels equals 60\arcsec{} on the solar disc in the linear scale. In one observational day, 80 pairs of velocity maps are calculated and averaged.

\section{Method calibration}
In our tests we have used lots of variations of simple axisymmetric model flows (with a wide range of values of parameters describing the differential rotation and meridional circulation) with good success in reproducing the models. When comparing the resulting vectors of motions with the model ones, we found a systematic offset in the zonal component equal to $v_{\rm offset,\ zonal}=-15\ \rm m\,s^{-1}$. This constant offset appeared in all the tested model velocity fields and comes from the numerical errors during the  ``derotation'' of the whole time series. For the final testing, we used one of the velocity fields obtained in our previous work \citep{svanda05}. This field approximates the velocity distribution that we may expect to observe on the Sun. The model flows have structures with a typical size of 60\arcsec{}, since they were obtained with the correlation window of this size.

\begin{figure}[!b]
\resizebox{0.50\textwidth}{!}{\includegraphics{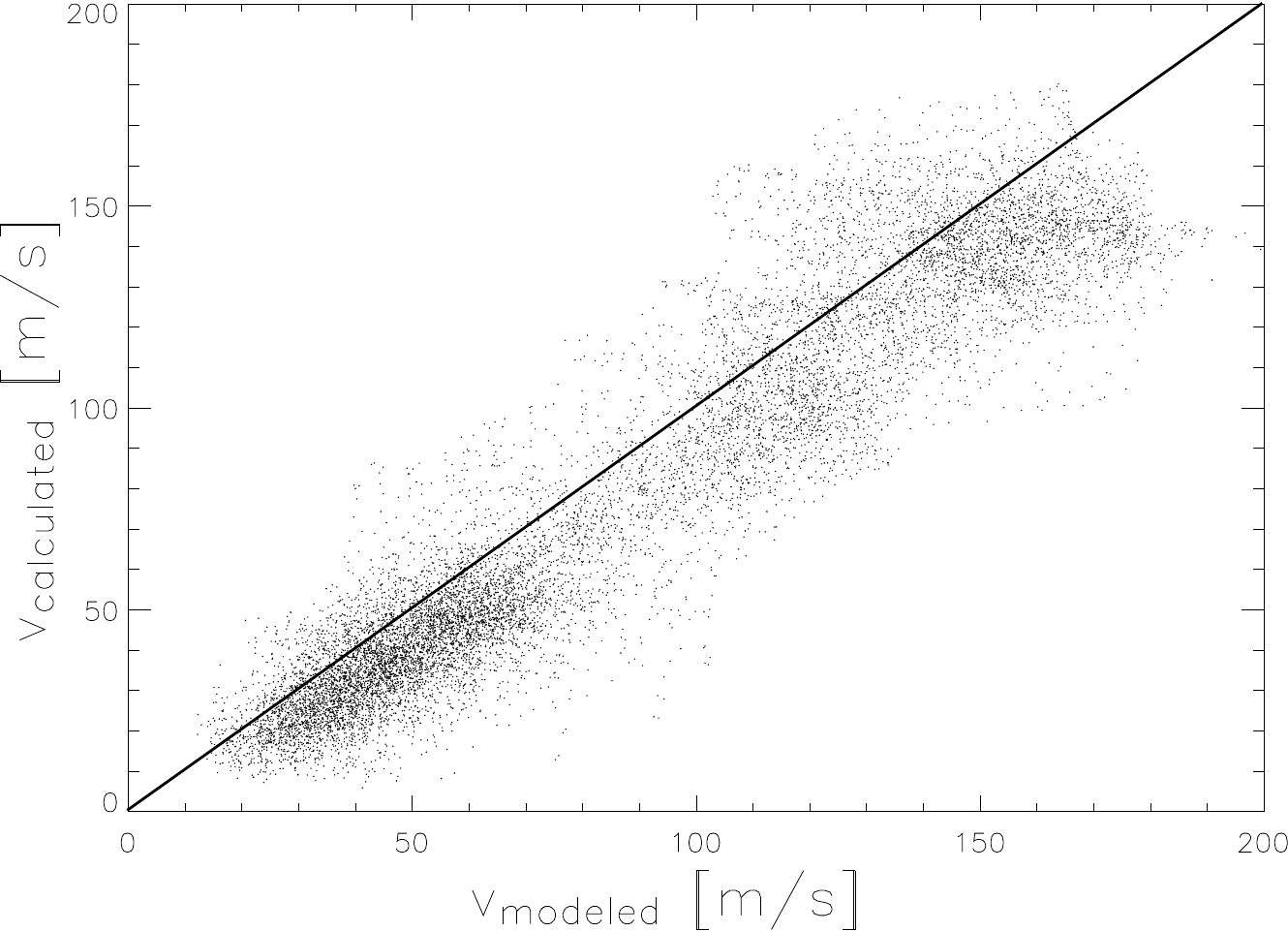}}
\caption{Scatter plot for inference of the calibration curve. Magnitudes of calculated velocities are slightly underestimated by LCT, but the linear behaviour is clearly visible. A line representing the 1:1 ratio is displayed. The calibration affects only the magnitudes of the flows, while the directions do not need any correction.}
\label{svanda_fig:kalibrace}
\end{figure}

The calculated velocities (with $v_{\rm offset,\ zonal}=-15\ \rm m\,s^{-1}$ corrected) were compared with the model velocities. Already from the visual impression it becomes clear that most of vectors are reproduced very well in the direction, but the magnitudes of the vectors are not reproduced so well. Moreover, it seems that the magnitudes of vectors are underestimated. This observation is confirmed when plotting the magnitudes of the model vectors versus the magnitudes of the calculated vectors (Fig.~\ref{svanda_fig:kalibrace}). The scatter plot contains more than 1 million points, and most of the points concentrate along a strong linear dependence, which is clearly visible. This dependence can be fitted by a straight line that can be used to derive the calibration curve of the magnitude of calculated velocity vectors. The calibration curve is given by the formula
\begin{equation}
v_{\rm cor}=1.13\,v_{\rm calc},
\label{svanda_eq:calibration}
\end{equation}
where $v_{\rm calc}$ is the magnitude of velocities coming from the LCT, and $v_{\rm cor}$ the corrected magnitude. The directions of the vectors before and after the correction are the same. The uncertainty of the fit can be described by 1-$\sigma$-error 15~m\,s$^{-1}$ for the velocity magnitudes under 100~m\,s$^{-1}$ and 25~m\,s$^{-1}$ for velocity magnitudes greater than 100~m\,s$^{-1}$. The uncertainties of approx. $15\ \rm m\,s^{-1}$ have their main origin in the evolution of supergranules. 

\section{Comparison to time-distance helioseismology}
Solar acoustic waves ($p$-modes) are excited in the upper convection
zone and travel between various points on the surface through the
interior. The travel time of acoustic waves is affected by
variations of the speed of sound along the propagation paths and
also by mass flows. Time-distance helioseismology measurements
\cite{1993Natur.362..430D} and inversions
provide (see references to Duvall's paper) a tool to study three-dimensional flow fields in the upper
part of the solar convection zone with relatively high spatial
resolution.

\begin{figure*}[!t]
\begin{center}
\resizebox{!}{11cm}{\includegraphics{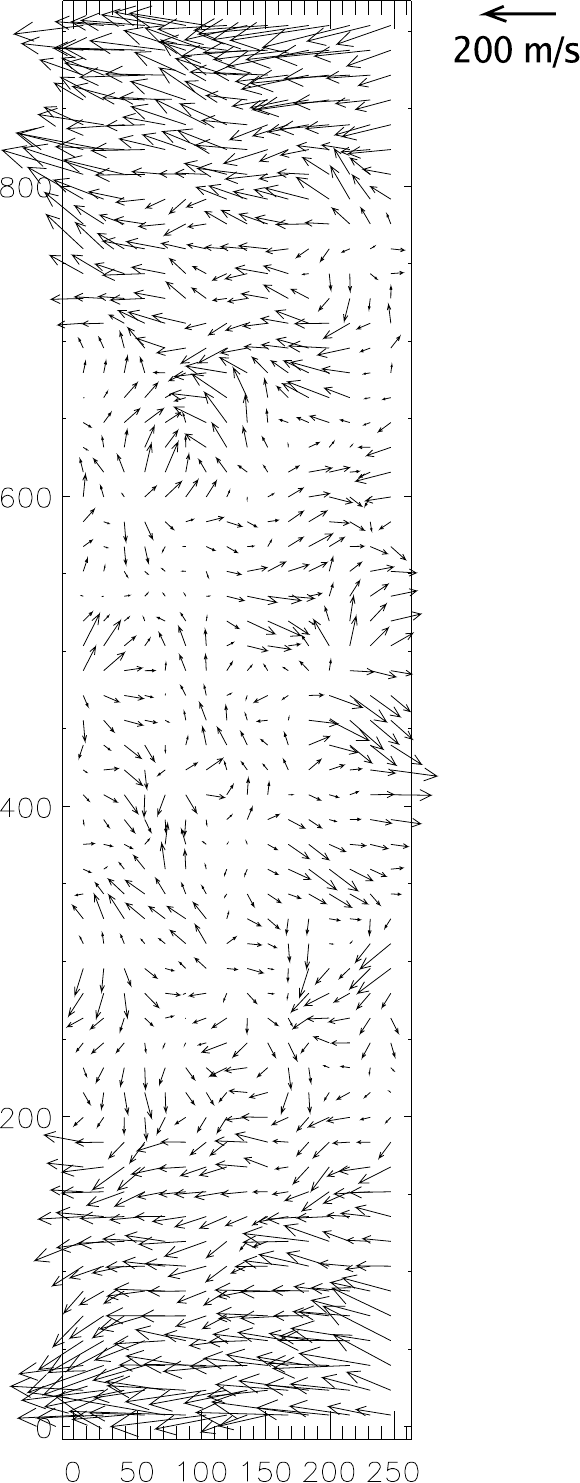}}
\resizebox{!}{11cm}{\includegraphics{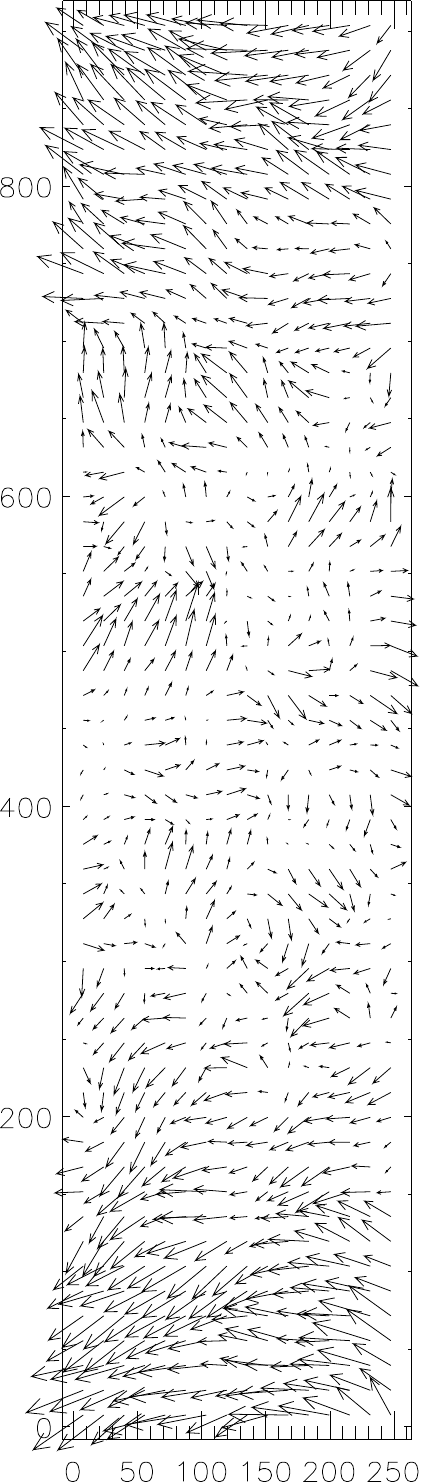}}
\caption{\emph{Left} --
Velocity field obtained by the LCT method. \emph{Right} -- Velocity
field obtained by the time-distance technique. Both plots are
centered at heliographic coordinates $b_0=0.0\,^\circ$,
$l_0=214.3\,^\circ$. Units on both axes are pixels in the data frame
with resolution of 0.12\,$^\circ$\,px$^{-1}$ in the Postel's
projection.} \label{arrows}
\end{center}
\end{figure*}
The local correlation tracking method and the method of time-distance helioseismology provide surface or near-surface velocity vector fields.
However, the results of these methods can be interpreted
differently. While local helioseismology measures intrinsic plasma
motions (through advection of acoustic waves), LCT measures apparent
motions of some structures (granules or magnetic elements). It is
known that some structures do not necessarily follow the flows of
the plasma on the surface. For example, the supergranulation appears to rotate
faster than the plasma \citep{2000SoPh..193..333B}. 
Some older studies \citep[see e.\,g.][]{1991BAAS...23.1033R} also suggest 
that the difference in flow properties measured on the basis of structures' motions and 
plasma motions is caused by a deeper anchor depth of these structures. An evolution of pattern 
may also play significant role (e.\,g. due to emergence of
magnetic elements).

Some attempts to compare the results of local helioseismology and the
LCT method for large scales, with characteristic size 100~Mm and
more, have been carried out by \cite{2005ASPC..346....3A}, but his results were inconclusive. The
correlation coefficient describing the match of the velocity maps
obtained by local helioseismology and the LCT method was close to
zero. Nevertheless, there were compact and continuous regions of the
characteristic size from 30 to 60 heliographic degrees with a good
agreement between the two methods, so that one could not conclude
that the results were completely different. In his study, many
factors could be significant: the techniques were applied to
different types of datasets (LCT was applied to low resolution
magnetograms acquired at Wilcox Solar Observatory and the
time-distance method used MDI Dopplergrams). Both techniques had
very different spatial resolution, and also the accuracy of the
measurements was not well known.

We decided to avoid these problems and analyze the same data set
from the MDI instrument on SOHO. MDI provides approximately two
months of continuous high-cadence (1 minute cadence) full-disc
Dopplergrams each year. This \emph{Dynamics Program} provides data
suitable for helioseismic studies, and also for the local correlation
tracking of supergranules. Thus, this is a perfect opportunity to
compare the performance and results of two different techniques using
the same set of data, and to avoid effects of observations with
different instruments or in different conditions.

The selected dataset consists of 27 data-cubes from March 12th,
2001, 0:00~UT to April 6th, 2001, 0:31~UT, where each third day was
used, and in these days three 8.5-hour long data-cubes were
processed (so that every third day in the described interval was
fully covered by measurements). Each data-cube is composed of 512
Dopplergrams with spatial resolution of 1.98$^{\prime\prime}\,\rm px^{-1}$ 
in one-minute cadence. All the frames of each data-cube were tracked with rigid
rate of 2.871~$\mu$rad\,s$^{-1}$, remapped to the Postel's
projection with resolution of 0.12\,$^\circ$\,px$^{-1}$ 
(1,500~km\,px$^{-1}$ at the center of the disc), 
and only a central meridian region was selected for the ongoing processing
(with size of 256$\times$924~px covering 30 heliographic degrees in
longitude and running from $-54\,^\circ$ to $+54\,^\circ$ in
latitude), so that effects of distortions due to the projection do
not play a significant role. The time-distance inversion results were smoothed by a Gaussian
with FWHM of 30~px to match the resolution to the LCT method, and
only the horizontal components ($v_x$, $v_y$) of the full velocity
vector were used.

The results containing 27 horizontal flow fields were statistically
processed to obtain the cross-calibration curves for these methods.
It is generally known 
that the LCT method slightly underestimates the
velocities; thus, the results should be corrected by a certain
factor. From the comparison of the $x$-component of velocity we obtained parameters of a linear fit given
by (numbers in parentheses denote a $\sigma$-error of the regression
coefficient)
\begin{equation}
v_{x,{~\rm LCT}}=0.895(0.008) v_{x,{~\rm t-d}}-12.6(0.3)~{\rm m\,s^{-1}} .
\label{vx_fit}
\end{equation}
The correlation coefficient between $v_{x,{~\rm LCT}}$ and
$v_{x,{~\rm t-d}}$ is $\rho=0.80$. We assume that the time-distance
measurements for $v_{x,{~\rm t-d}}$ are correct and the magnitude
of the LCT measurements, $v_{x,{~\rm LCT}}$, must be corrected
according to the slope of Eq.~(\ref{vx_fit}). This correction factor
has a value of 1.12, which is in a perfect agreement with the
correction factor of 1.13 found in the tests of the same LCT code
using synthetic Dopplergrams with the same resolution and similar
LCT parameters. We assume that both velocity
components obtained with the LCT method should be corrected by this
factor. The regression line of $v_y$ component is
\begin{equation}
v_{y,{~\rm LCT}}=0.56(0.01) v_{y,{~\rm t-d}}+0.4(0.2)~{\rm m\,s^{-1}} .
\label{vy_fit}
\end{equation}
After the slope correction using the $v_x$ fits, the regression
curve is slightly different:
\begin{equation}
v_{y,{~\rm LCT}}=0.63(0.01) v_{y,{~\rm t-d}}+0.4(0.2)~{\rm m\,s^{-1}},
\end{equation}
with the correlation coefficient between $v_{y,{~\rm LCT}}$ and
$v_{y,{~\rm t-d}}$ close to 0.47. The slope of the linear fit
differs significantly from the expected value 1.0. 

We have tested that this asymmetry is not related to the LCT
technique. The tests did not show any preference in direction of flows
measured by LCT or any dependence of the results on the size
of the field of view (which does not have a square shape in our case). In our synthetic data based test we did not encounter asymmetry of any kind.

Instead of computation of the correlation coefficient of the arguments 
of both vector fields we decided to compute a magnitude-weighted cosine 
of $\Delta \varphi$. This quantity is given by 
\begin{equation}
\rho_{\rm W} = \frac{\sum |{\mathbf v}_{t-d}| \frac{{\mathbf v}_{t-d} \cdot {\mathbf v}_{LCT}}
{|{\mathbf v}_{t-d}||{\mathbf v}_{LCT}|}}{\sum |{\mathbf v}_{t-d}|} = \frac{\sum |{\mathbf v}_{t-d}| \cos \Delta\varphi}{\sum |{\mathbf v}_{t-d}|}, 
\end{equation}
where ${\mathbf v}_{t-d}$ is the time-distance vector field, ${\mathbf v}_{LCT}$ is the LCT vector field
and the summation is performed over all vectors in the field. The closer this quantity is to 1, the 
better is the agreement between both vector fields. Larger vectors are weighted more than smaller ones. 
We have found that in our case $\rho_{\rm W}=0.86$.

In addition to the detailed comparison of the vector fields, we
compare the mean flows, the differential rotation and the meridional
circulation. These flows can be quite simply calculated from the
results of both techniques. In both cases, they provide the mean
zonal and mean meridional flows for the Carrington rotation
No.~1974. The correlation coefficients are $\rho=0.98$ for the
zonal flow and $\rho=0.88$ for the meridional flow. 

For a detailed comparison of the flow fields, we selected one data
cube, representing 8.5-hour measurements centered at 4:16~UT of March
24th, 2001, $l_0$=214.3\,$^\circ$. In this map, the
correlation coefficient for the $x$-component of the velocity is
$\rho=0.82$, for the $y$-component $\rho=0.58$, and for the
vector magnitude: $\rho=0.73$. The vector plots of the flow fields
obtained by both techniques, shown in Fig.~\ref{arrows}, in general seem to be
quite similar to each other. However, many differences can be seen. The
regions  where the differences are most significant correspond to
relatively small (under 50~m\,s$^{-1}$) velocities.

\section{Application to real data}
\subsection{Long-term properties}
For the study of long-term evolution of surface flows, maps containing the mean zonal and meridional components were calculated. We can clearly detect that in the northern hemisphere the flow towards the northern pole dominates while in the southern hemisphere the flow towards southern pole prevails. The ``zero line'', the boundary between the flow polarity, is not located exactly on the solar equator and seems to be shifted to the south in the period of increased solar activity (2001 and 2002). In agreement with \cite{1997AA...319..683M} and \cite{1998SoPh..183..263C} we found the meridional flow stronger in the periods of increased solar activity by about $~10$~\mps{} than in periods with lower magnetic activity.

\begin{figure*}
\centering
\includegraphics[width=0.9\textwidth]{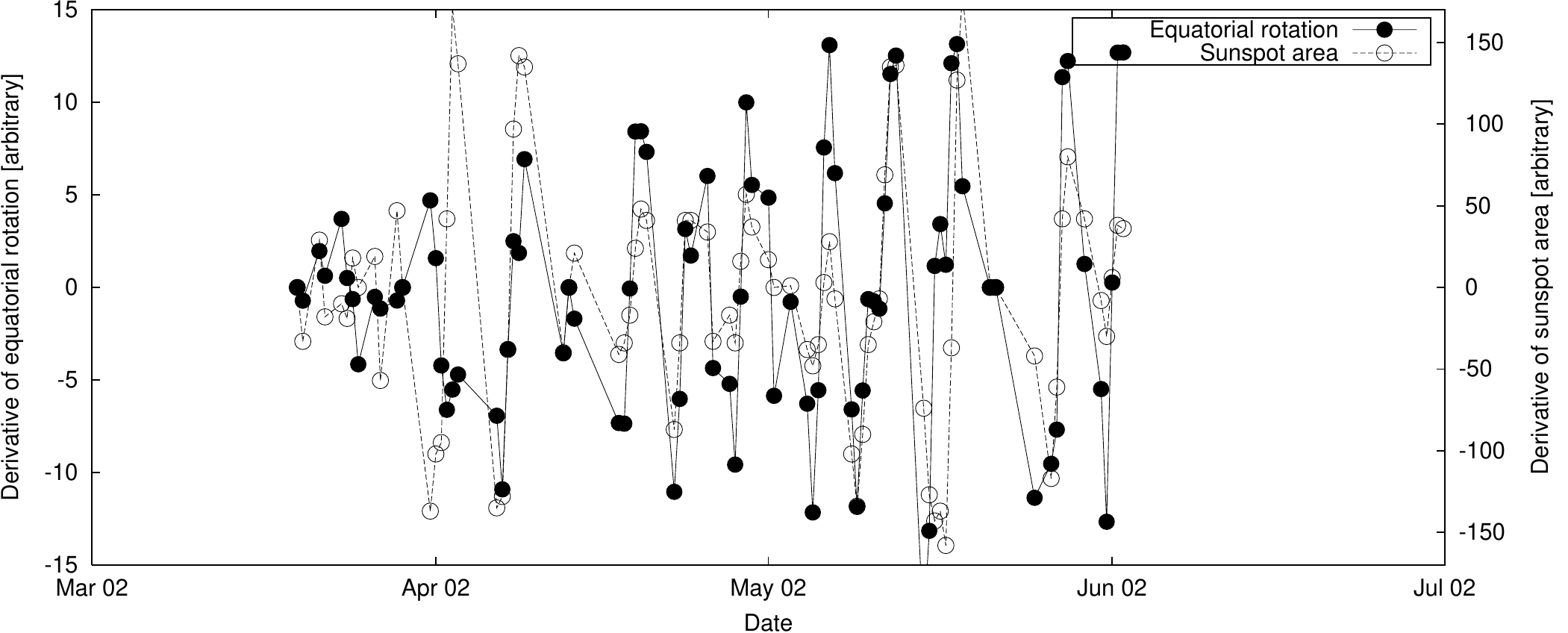}
\caption{Derivatives of the mean zonal velocity (solid curve) and the sunspot area in the near-equatorial region (dashed curve) in 2002. Both quantities correlate with each other well.}
\label{fig:gradients}
\end{figure*}
A similar map was made in the same way for the zonal component. The mean equatorial zonal velocity for all the data is 1900~\mps. For all the processed data the dependence on latitude is close to a parabolic shape, the parameters of which change slowly over time. The residua of the zonal velocity with respect to its parabolic fit given by
\begin{equation}
v_b=a_0+a_1 b + a_2 b^2,
\end{equation}
where $b$ is the heliographic latitude and $v_b$ the mean zonal velocity in the given latitude, were calculated in
order to see if we are able detect torsional oscillations in our measurements. The method reveals torsional oscillations as an excess of the mean zonal velocity with respect to the zonal velocity in the neighbourhood. The behaviour of torsional oscillations is in agreement with their usual description -- the excess in magnitude is of the order of 10~\mps, they start at the beginning of the solar cycle in high latitudes and propagate towards the equator with the progress of the 11-year cycle. However, with our method the visibility of torsional oscillations decreases with increasing solar activity. In the periods of strong activity both belts are not so clearly visible since the large-scale velocity field and its parabolic fit are strongly influenced by the presence of magnetic regions. However, the torsional oscillation belts still remain visible when the mean zonal component is symmetrised with respect to the solar equator. We did not focus on the study of meridional flow or torsional oscillations depending on time and latitude, we just used them to check the ability and performance of our method.

We investigated the coupling of equatorial zonal velocity (average equatorial solar rotation) and the solar activity in the near-equatorial area (belt of heliographic latitudes from $-10\,^\circ$ to $+10\,^\circ$). The average equatorial zonal velocity incorporates the average supergranular network rotation and also the movement of degenerated supergranules influenced by a local magnetic field with respect to their non-magnetic vicinity. Indexes of the solar activity were extracted from the daily reports made by the \emph{Space Environment Center National Oceanic and Atmospheric Administration (SEC NOAA)}. Only the days when the measurements of horizontal flows are available were taken into account. As the index of the activity we have considered the total area of sunspots in the near-equatorial belt and also their type.

First, we computed the correlation coefficient $\rho$ between the mean equatorial zonal velocity and the sunspot area in the near-equatorial belt and obtained a value of $\rho=-0.17$. We cannot conclude that there is a linear relation between these two indices. We find two different regimes which are divided by the velocity of approximately 1890~\mps. In one regime (77~\% of the cases), the equatorial belt rotates about 60~\mps{} faster ($1910 \pm 9$~\mps; hereafter the ``fast group'') than the Carrington rotation, in the other one (23~\%) the rotation rate is scattered around the Carrington rate ($1860 \pm 20$~\mps; hereafter the ``scattered group''). The division in two groups using the speed criterion is arbitrary. If there are only two groups, they certainly overlap and only a very detail study could resolve their members. The arguments for division in just two groups will follow.

Detailed studies of the sunspot drawings obtained from the Patrol Service of the Ond\v{r}ejov Observatory and the Mt.~Wilson Observatory drawings archive revealed that in the ``fast'' group, the new or growing young active regions were present in the equatorial belt. On the contrary, in the ``scattered'' group the decaying or recurrent active regions prevailed in the equatorial area. The deceleration of the sunspot group with its evolution was noticed e.\,g. by \cite{2004SoPh..221..225R}. Moreover, our results suggest that the new and rapidly growing sunspots in the studied sample (March to May 2001 and April to June 2002) move with the same velocity. This behaviour could be explained by emergence of the local magnetic field from a confined subphotospheric layer. According to the rough estimate \citep{2002SoPh..205..211C} the speed of $1910 \pm 9$~\mps{} corresponds to the layer of $0.946 \pm 0.008\ R_\odot$, where the angular velocity of rotation suddenly changes. During the evolution, the magnetic field is disrupted by the convective motions. 

The observed behaviour could be a manifestation of the disconnection of magnetic field lines from the base of the surface shear during the evolution of the growing sunspot group. This behaviour was studied theoretically by \cite{2005AA...441..337S}. They suggested a dynamic disconnection of bipolar sunspot groups from their magnetic roots deep in the convection zone by upflow motions within three days after the emergence of the new sunspot group. The motion of sunspots changes during those three days from ``active'' to ``passive''. The active mode is displayed by motions that are faster than those of a non-magnetic origin. The passive mode means mostly the deceleration of sunspot motion and influence of the sunspot motion only by the shallow surface plasma dynamics. The theory of the disconnection of sunspot groups from their magnetic root supports the division of the data set in two groups. 

We have also focused on how the presence of the magnetic active areas will influence the average flow field. Since we found that the direct correlation is weak due to the existence of two different regimes, we decided to study the temporal change of both quantities. Our aim is to study whether an emerging active region in the near-equatorial belt will influence the average equatorial rotation. We computed numerical derivatives of the total sunspot area in the near-equatorial belt and of the average zonal equatorial flow. We found that the correlation coefficient between both data series is $\rho=0.36$ and is higher for the ``fast group'' ($\rho=0.41$) than for the ``scattered group'' ($\rho=0.24$). The correlation is higher in periods of increased magnetic activity in the equatorial belt. For example, for data in year 2001 the correlation coefficient is $\rho_{2001}=0.58$ and for year 2002 $\rho_{2002}=0.52$ (see Fig.~\ref{fig:gradients}). In both cases, the correlation is higher for the ``fast regime'' ($\rho \sim 0.7$) than for the second group. 

\begin{figure*}
\centering
\includegraphics[width=0.45\textwidth]{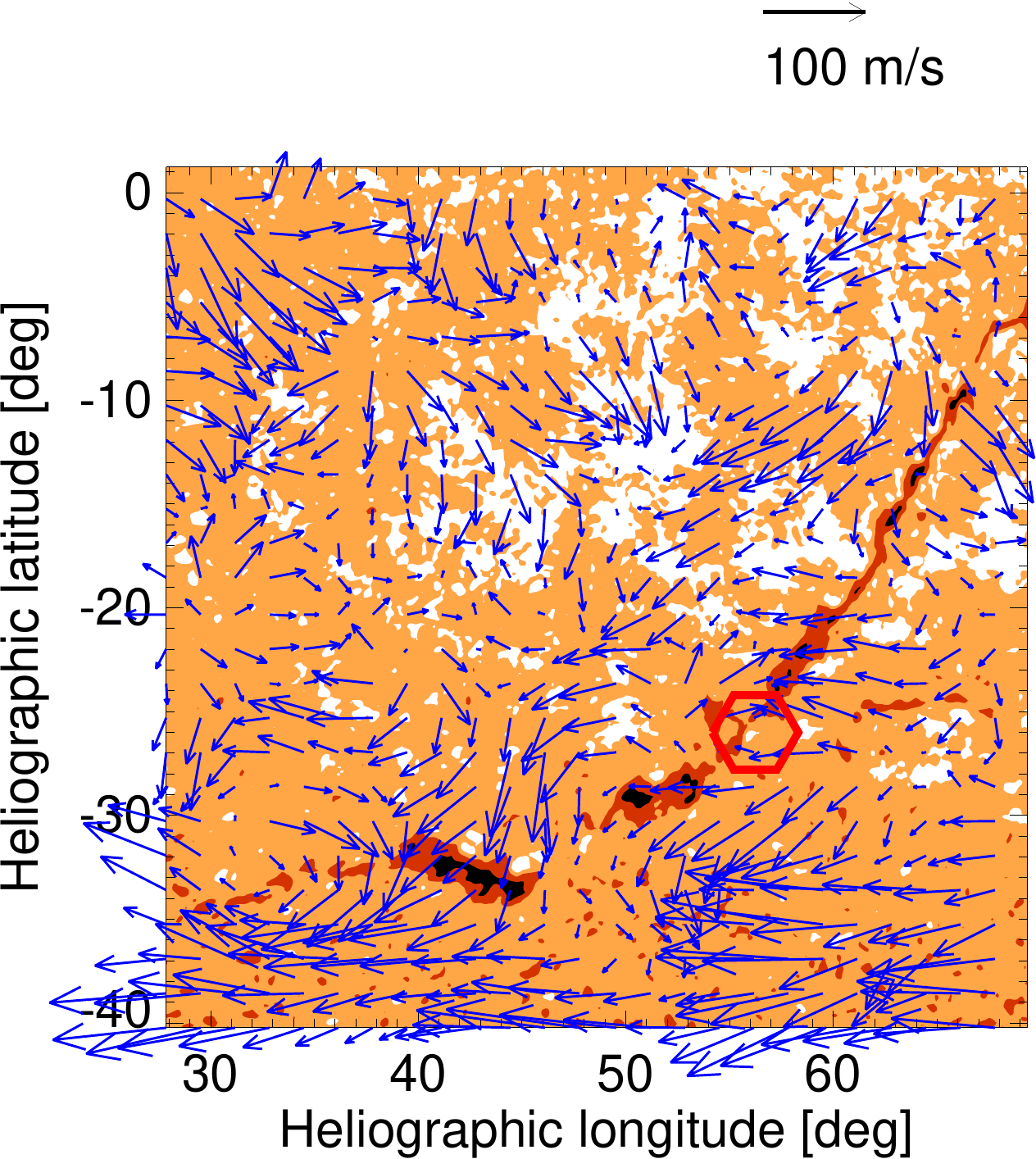}
\includegraphics[width=0.45\textwidth]{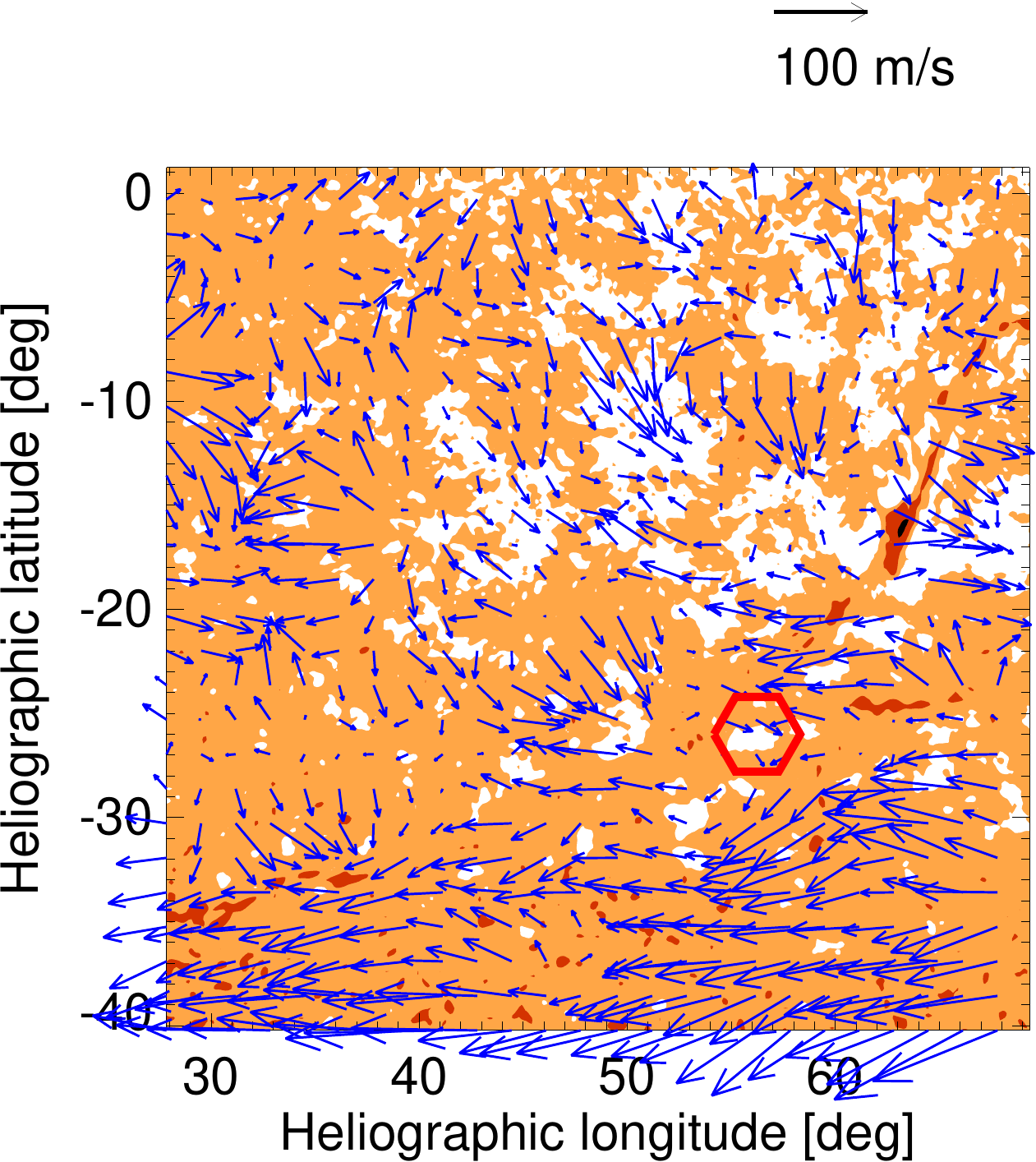}
\caption{The 3-hour average of the flow field in the close vicinity of the starting point ($l=56^\circ$, $b=-26^\circ$)
before (left) and after (right) filament eruption. The filament observed by ISOON on October 7 2004  at 15.20~UT is superimposed. The dashed boxes denote areas used for the zonal shear calculation.}
\label{context6}
\end{figure*}
Despite the apparent disagreement, our seemingly conflicting results can be valid and are in agreement with results published earlier. As described by e.\,g. \cite{1990ApJ...357..271H} and explained in the model of \cite{2004SoPh..220..333B}, the solar rotation in the lower latitudes is slower in the presence of a magnetic field. In most cases, ``spotty'' equatorial belts seem to rotate slower than average for the whole data series. However, it is clear that in most cases emerging active regions cause an increase in the rotation rate. This is in agreement with the generally accepted findings of \cite{1970SoPh...12...23H} and \cite{1978ApJ...219L..55G}. The relationship, obtained using a linear fit on our data set, can be described by the equation
\begin{equation}
\Delta v \sim 0.2\, \Delta A_{\rm sunspots}\ {\rm m\,s^{-1}},
\end{equation}
where $\Delta v$ is the change of the equatorial rotation speed with respect to the Carrington rotation and $\Delta A_{\rm sunspots}$ is the relative change of sunspot area in the equatorial belt (in 10$^{-6}$ of solar hemisphere). We estimate that strong local magnetic areas rotate a few tens of \mps{} faster than the non-magnetic surroundings.

\subsection{Flows around and beneath the erupting filament}
During three consecutive days of the JOP 178 campaign, Oct 6, 7, and 8, 2004, we observed the evolution of a filament that was close to the central meridian. We also observed the photospheric flows directly below the filament and in its immediate area. 
The filament extends from $-$5\degr{} to $-$30\degr{} in latitude. A filament eruption
was observed on October 7, 2004 at 16:30~UT at   multiple wavelengths from ground and space instruments.

At the point where the filament eruption begins ($l=56^\circ$, $b=-26^\circ$ in Carrington coordinates),
we detected a steepening of the gradient in the differential rotation curve. During the eruption, the gradient
flattens out and a dip forms. Although the differential rotation curves describe
the mean zonal velocities on the full-disc, the change of its gradient signify the change in the stretch
influencing the magnetic field in the loop over the area under study. The parameters of 
the smooth fitted curve did not change too much, however, the local residual with respect to the smooth curve changed at the 
latitude where the filament eruption starts.

Fig.~\ref{context6} displays the horizontal flows over a wide field of view measured using the
our supergranule-tracking method and averaged over 3~hours, before and after the 
filament eruption. Before the eruption we can clearly see the north--south stream parallel to and about  10\degr{} east
of the filament. This stream disturbs differential rotation and brings plasma and magnetic structures to the south. Although 
differential rotation tends to spread the magnetic lines to the east, the observed north--south stream
tends to shear the magnetic lines. After the eruption, only a northern segment of the filament is visible and
the north--south stream has disappeared.

We measured the shear as a difference between the mean zonal component $v_x$ in the area. We measured the average zonal flow over boxes 2.3\degr{} on a side located 2.9\degr{} north and south of the point where the filament eruption appeared to start. Six 2-hour averaged of the flow fields were used to determine this figure. One can see that the shear velocity is increasing before the eruption and decreasing after the eruption. One hour before the eruption the shear reached the value of (120$\pm$15)~m\,s$^{-1}$ over a distance of 5.2\degr{} (62\,000~km in the photosphere). After the filament eruption we observe ordinary differential rotation restoration below 30\degr{} south.

\subsection{Vertical structure of horizontal flows}

The issue of the our tracking-based method is that we measure displacements of supergranular structures in the series of processed Dopplergrams and interpret them as the large-scale velocity field in the solar photosphere. However, we cannot establish the range of depths in the solar convection zone, where the large-scale flows are well represented by the surface measurements. With the use of the data provided by the time-distance helioseismology we can determine this unknown parameter. We can also verify the assumption of our analysis: whether supergranules are subjected to the transport by the velocity field on the larger scale.

The large-scale flows are a combination of many types of motions of various spatial scales (such as differential rotation, meridional circulation, possibly giant cell convection, and others), which unfortunately cannot be reliably separated in components. These components may vary independently with time and depth. This implies that, in principle, when dealing with the large-scale flows, we cannot obtain unambigous results that do not allow any alternative interpretation. Nonetheless, the realistic numerical simulation should reveal the combined large-scale plasma flow, properties of which is comparable with the measured ones. Such numerical simulation, which is not present at the time, shall allow to distinguish various components of the detected large-scale flow and to study them separately. The results therefore provide encouraging large-scale flow properties, which are to be reproduced by models.

For the comparison between the data obtained by our method and the results of the time-distance helioseismology we used high-cadence full-resolution full-disc Dopplergrams recorded in March and April 2001. In this period, 46 flow maps averaged over 8~hours of the vicinity of active region NOAA~9361/9393/9433 were processed by time-distance helioseismology using the ray approximation \citep[see][]{2004ApJ...603..776Z}. From these maps we used six examples in six different days (March 3rd, March 28th, 29th, 30th, 31st, and April 25th). In these days, MDI data series did not contain many gaps, so the data were also suitable for our tracking method, and the field of view was far from the solar limb, so we can exclude any possible effect of undersampling in Dopplergrams near the solar limb.

We had to slightly modify our method to match challenges and properties of the time-distance data. Postel projection was used instead of Sanson-Flamsteed ones, because helioseismic data are computed in this geometrical projection (it conserves the main circles, therefore it is suitable for measurement of $p$-modes' travel-times).

All the datasets were aligned with the centre of the field of view on Carrington coordinates $l=148.5^\circ$ and $b=19^\circ$. The field of view has the total size of 512$\times$512 pixels with the resolution of 0.92\arcsec{}\,px$^{-1}$. The datacubes from the local helioseismology contain 15 irregularly spaced depths from 0.77~Mm to more than 80~Mm. The results of our method have effective resolution of 60\arcsec{}, but the results of time-distance helioseismology have 8\arcsec. We smoothed helioseismic flow maps to match the resolution of 60\arcsec. This also means that the structure of the internal flow in supergranulation is filtered out and does not disturb the study of the large-scale flow field, which is the most important deviation from the study of \cite{2003ESASP.517..417Z}. The velocity field obtained by our tracking method are in principle twodimensional -- horizontal. The $z$-component of the time-distance flows suffers from effects of cross-talks, therefore we use only horizontal (i.\,e. in $x$ and $y$) components of the time-distance flows. For each depth in the helioseismic datacube we calculated its similarity to the flow map obtained by tracking of supergranular structures. As the measure of similarity we used the magnitude weighted cosine of the direction difference.

\begin{figure}[!t]
\includegraphics[width=0.5\textwidth]{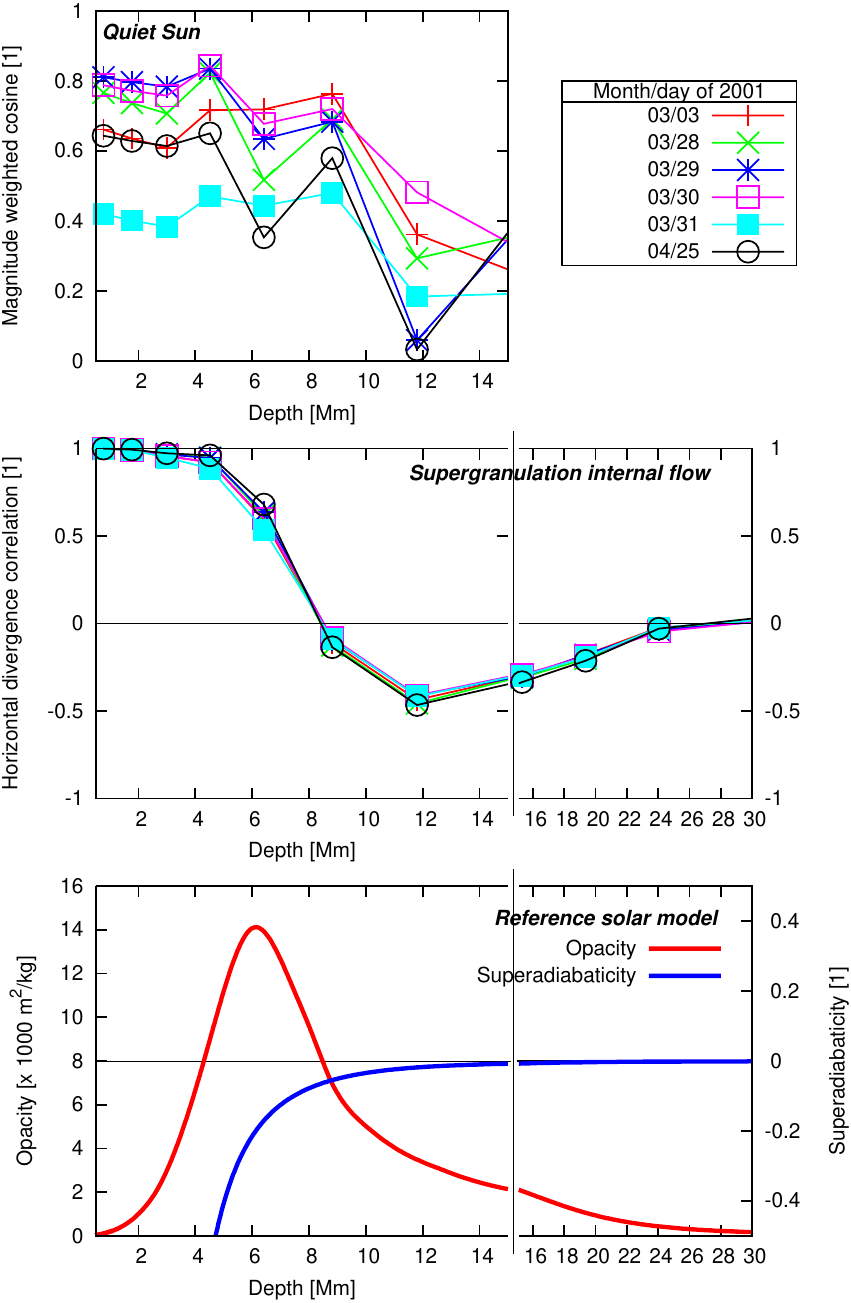}
\caption{\emph{Upper:} The similarity of the large-scale flows at various depths in the solar convection zone with the surface measurements by LCT. \emph{Middle:} Correlation of the divergence signal representing supergranulation between the small-scale flow maps at various dephts and the surface layers. \emph{Bottom:} Trends of the opacity and superadiabaticity according to reference model S in the same level in the solar convection zone.}
\label{fig:vertical}
\end{figure}

Our analysis shows that the large-scale flow does not vary much with depth down to 10~Mm in the quiet Sun regions (see Fig.~\ref{fig:vertical}). This fact means that the large-scale horizontal velocity field measured in the surface measurement represents well the large-scale horizontal dynamics in the layer of 0--10~Mm depth. The obtained results can be interpreted as a fulfilment of the basic assumption of our tracking method that supergranular structures may be indeed treated as the objects carried by the large-scale velocity field and that the supergranules move in the depths down to 5~Mm coherently in the horizontal direction with a slight lost of coherence between 5 and 10~Mm. 

There is an alternative interpretation of the obtained results. Assuming that the time-distance helioseismology is less reliable deeper down in the convection zone and that the large-scale flows do not vary much in the entire upper convection zone, our results put a level, where the reliability of the time-distance measurements is reasonable. In this interpretation, the time-distance flow measurements are reasonable in upper 10~Mm in the quiet Sun regions and upper 5~Mm in regions occupied by the magnetic field. There are recent papers \citep[e.\,g.][]{2008SoPh..251..381J} using another time-distance implementation, which suggest the range of reasonability of time-distance results only upper 3~Mm. 

To confirm our first interpretation, the structure of the supergranules itself must be analysed. Therefore we modified our analysis and instead of smearing the time-distance flow field to remove the signal of supergranulation internal mass flow, we subtracted the large-scale velocity field, so that only the internal velocities in supergranules remained in the flow map. From the maps containing only the separated supergranulation signal we calculated the horizontal divergence as the representation of down/up-flows. It is more stable than the measured $z$-component of velocity, which suffers of plenty from issues \citep{Zhao2006priv}. 

In the middle panel in Fig.~\ref{fig:vertical} we see that the divergence signal is very similar in the quiet Sun region in depths 0.77--6.4~Mm. Down to $\sim$7~Mm the correlation is positive, implying the coherency in the structure of supergranules. Deeper down the correlation turns negative, which might suggest the evidence of return flow in supergranules. 

In the lowest panel in Fig.~\ref{fig:vertical} the trend of the superadiabaticity within the sub-surface layers of convection zone, coming from reference solar model~S \citep{1996Sci...272.1286C} is plotted. The superadiabaticity $A^*$ is defined as
\begin{equation}
A^*=\frac{1}{\gamma} \frac{{\rm d} \ln p}{{\rm d} \ln r}-\frac{{\rm d} \ln \rho}{{\rm d} \ln r}\ ,
\end{equation}
where $\gamma$, $p$, and $\rho$ are the state parametres (adiabatic exponent, pressure, and density) of the plasma at distance $r$ from the center of the Sun. Should the $A^*$ be negative, the layers at $r$ are convectively unstable \citep[e.\,g. discussion in][]{1997ApJ...474..790D}. It is to be noticed that $A^*$ turns more negative at the depths of ~10--12~Mm, where it is assumed that the supergranulation should be formed, and remain very negative up to the solar photosphere, where it turns positive again (the photosphere is convectively stable). It is the same layer, where the large-scale flow is coherent. It may be interpreted that the supergranules that form at depth some 10~Mm below the surface are carried by the large-scale velocity field, which operates here. Therefore the large-scale flow remain nearly constant within upper 10~Mm of the convection zone and is detected using our tracking method in surface measurements. 

In the same panel the trend of opacity, comming also from model S, is overplotted. Following the old opacity concept of conditions for onset of the convection \citep{1975ApJ...195..137S}, we may conclude that from this point of view, there are suitable physical conditions at the depths of 10~Mm to form the convection pattern of supergranulation.

\section{Conclusions}

\begin{enumerate}
\item Tracking of supergranules makes a perfect sense when studying the large-scale field in the solar photosphere and near subphotospherical layer.
\item Our method provides results with effective resolution of 60\arcsec{}. 
\item Standard LCT lag between correlated frames is 4 hours, but is changeable, when averaging over 24 hours, the noise level is at 15 m/s (random error), other spans untested.
\item SOHO/MDI provides enough material for some other studies, HMI/SDO will do better.
\item We demonstrate the power of this method on a few case studies. We intend to continue in this research and expect the new results describing the amazing dynamics in the upper solar convection zone.
\end{enumerate}

\section*{Acknowledgements}
M.~\v{S}, M.~K., and M.~S. were supported by the Grant Agency of Academy of Sciences of the Czech Republic under grant IAA30030808, M.~\v{S} additionally by ESA-PECS under grant 98030. The Astronomical Institute of ASCR is working on the Research project AV0Z10030501 (Academy of Sciences of CR), the Astronomical Institute of Charles University on the Research program MSM0021620860 (Ministry of Education of CR). SOHO is a project of international cooperation between ESA and NASA.

\begin{thebibliography}{31}
\expandafter\ifx\csname natexlab\endcsname\relax\def\natexlab#1{#1}\fi

\bibitem[{{Ambro{\v z}}(2001{\natexlab{a}})}]{2001SoPh..198..253A}
{Ambro{\v z}}, P. 2001{\natexlab{a}}, \solphys, 198, 253

\bibitem[{{Ambro{\v z}}(2001{\natexlab{b}})}]{2001SoPh..199..251A}
{Ambro{\v z}}, P. 2001{\natexlab{b}}, \solphys, 199, 251

\bibitem[{{Ambro{\v z}}(2005)}]{2005ASPC..346....3A}
{Ambro{\v z}}, P. 2005, in ASP Conf. Ser. 346: Large-scale Structures and their
  Role in Solar Activity, ed. K.~{Sankarasubramanian}, M.~{Penn}, \&
  A.~{Pevtsov}, 3

\bibitem[{{Beck} \& {Schou}(2000)}]{2000SoPh..193..333B}
{Beck}, J.~G. \& {Schou}, J. 2000, \solphys, 193, 333

\bibitem[{{Brun}(2004)}]{2004SoPh..220..333B}
{Brun}, A.~S. 2004, \solphys, 220, 333

\bibitem[{{Cameron} \& {Hopkins}(1998)}]{1998SoPh..183..263C}
{Cameron}, R. \& {Hopkins}, A. 1998, \solphys, 183, 263

\bibitem[{{Carrington}(1859)}]{1859MNRAS..19...81C}
{Carrington}, R.~C. 1859, \mnras, 19, 81

\bibitem[{{Christensen-Dalsgaard} {et~al.}(1996){Christensen-Dalsgaard},
  {Dappen}, {Ajukov}, {Anderson}, {Antia}, {Basu}, {Baturin}, {Berthomieu},
  {Chaboyer}, {Chitre}, {Cox}, {Demarque}, {Donatowicz}, {Dziembowski},
  {Gabriel}, {Gough}, {Guenther}, {Guzik}, {Harvey}, {Hill}, {Houdek},
  {Iglesias}, {Kosovichev}, {Leibacher}, {Morel}, {Proffitt}, {Provost},
  {Reiter}, {Rhodes}, {Rogers}, {Roxburgh}, {Thompson}, \&
  {Ulrich}}]{1996Sci...272.1286C}
{Christensen-Dalsgaard}, J., {Dappen}, W., {Ajukov}, S.~V., {et~al.} 1996,
  Science, 272, 1286

\bibitem[{{Corbard} \& {Thompson}(2002)}]{2002SoPh..205..211C}
{Corbard}, T. \& {Thompson}, M.~J. 2002, \solphys, 205, 211

\bibitem[{{Demarque} {et~al.}(1997){Demarque}, {Guenther}, \&
  {Kim}}]{1997ApJ...474..790D}
{Demarque}, P., {Guenther}, D.~B., \& {Kim}, Y.-C. 1997, \apj, 474, 790

\bibitem[{{Duvall} {et~al.}(1993){Duvall}, {Jefferies}, {Harvey}, \&
  {Pomerantz}}]{1993Natur.362..430D}
{Duvall}, Jr., T.~L., {Jefferies}, S.~M., {Harvey}, J.~W., \& {Pomerantz},
  M.~A. 1993, \nat, 362, 430

\bibitem[{{Golub} \& {Vaiana}(1978)}]{1978ApJ...219L..55G}
{Golub}, L. \& {Vaiana}, G.~S. 1978, \apjl, 219, L55

\bibitem[{{Hart}(1954)}]{1954MNRAS.114...17H}
{Hart}, A.~B. 1954, \mnras, 114, 17

\bibitem[{{Hathaway} \& {Wilson}(1990)}]{1990ApJ...357..271H}
{Hathaway}, D.~H. \& {Wilson}, R.~M. 1990, \apj, 357, 271

\bibitem[{{Howard} \& {Harvey}(1970)}]{1970SoPh...12...23H}
{Howard}, R. \& {Harvey}, J. 1970, \solphys, 12, 23

\bibitem[{{Jackiewicz} {et~al.}(2008){Jackiewicz}, {Gizon}, \&
  {Birch}}]{2008SoPh..251..381J}
{Jackiewicz}, J., {Gizon}, L., \& {Birch}, A.~C. 2008, \solphys, 251, 381

\bibitem[{{Leighton} {et~al.}(1962){Leighton}, {Noyes}, \&
  {Simon}}]{1962ApJ...135..474L}
{Leighton}, R.~B., {Noyes}, R.~W., \& {Simon}, G.~W. 1962, \apj, 135, 474

\bibitem[{{Meunier} {et~al.}(1997){Meunier}, {Nesme-Ribes}, \&
  {Collin}}]{1997AA...319..683M}
{Meunier}, N., {Nesme-Ribes}, E., \& {Collin}, B. 1997, \aap, 319, 683

\bibitem[{{November}(1986)}]{1986ApOpt..25..392N}
{November}, L.~J. 1986, \ao, 25, 392

\bibitem[{{November} \& {Simon}(1988)}]{1988ApJ...333..427N}
{November}, L.~J. \& {Simon}, G.~W. 1988, \apj, 333, 427

\bibitem[{{Rhodes} {et~al.}(1991){Rhodes}, {Korzennik}, {Hathaway}, \&
  {Cacciani}}]{1991BAAS...23.1033R}
{Rhodes}, Jr., E.~J., {Korzennik}, S.~G., {Hathaway}, D.~H., \& {Cacciani}, A.
  1991, in Bulletin of the American Astronomical Society, Vol.~23, Bulletin of
  the American Astronomical Society, 1033

\bibitem[{{Ru{\v z}djak} {et~al.}(2004){Ru{\v z}djak}, {Ru{\v z}djak}, {Braj{\v
  s}a}, \& {W{\"o}hl}}]{2004SoPh..221..225R}
{Ru{\v z}djak}, D., {Ru{\v z}djak}, V., {Braj{\v s}a}, R., \& {W{\"o}hl}, H.
  2004, \solphys, 221, 225

\bibitem[{{Schr\"oter}(1985)}]{1985SoPh..100..141S}
{Schr\"oter}, E.~H. 1985, \solphys, 100, 141

\bibitem[{{Sch{\"u}ssler} \& {Rempel}(2005)}]{2005AA...441..337S}
{Sch{\"u}ssler}, M. \& {Rempel}, M. 2005, \aap, 441, 337

\bibitem[{{Schwarzschild}(1975)}]{1975ApJ...195..137S}
{Schwarzschild}, M. 1975, \apj, 195, 137

\bibitem[{{{\v S}vanda} {et~al.}(2005){{\v S}vanda}, {Klva{\v n}a}, \&
  {Sobotka}}]{svanda05}
{{\v S}vanda}, M., {Klva{\v n}a}, M., \& {Sobotka}, M. 2005, Hvar Observatory
  Bulletin, 29, 39

\bibitem[{{{\v S}vanda} {et~al.}(2006){{\v S}vanda}, {Klva{\v n}a}, \&
  {Sobotka}}]{2006AA...458..301S}
{{\v S}vanda}, M., {Klva{\v n}a}, M., \& {Sobotka}, M. 2006, \aap, 458, 301

\bibitem[{{Zhao}(2006)}]{Zhao2006priv}
{Zhao}, J. 2006, {private communication}

\bibitem[{{Zhao} \& {Kosovichev}(2003)}]{2003ESASP.517..417Z}
{Zhao}, J. \& {Kosovichev}, A.~G. 2003, in ESA Special Publication, Vol. 517,
  GONG+ 2002. Local and Global Helioseismology: the Present and Future, ed.
  H.~{Sawaya-Lacoste}, 417

\bibitem[{{Zhao} \& {Kosovichev}(2004)}]{2004ApJ...603..776Z}
{Zhao}, J. \& {Kosovichev}, A.~G. 2004, \apj, 603, 776

\bibitem[{{Zhao} {et~al.}(2001){Zhao}, {Kosovichev}, \&
  {Duvall}}]{2001ApJ...557..384Z}
{Zhao}, J., {Kosovichev}, A.~G., \& {Duvall}, Jr., T.~L. 2001, \apj, 557, 384

\end{thebibliography}

\newcommand{\SortNoop}[1]{}

\end{document}